# Breast MRI radiomics and machine learning radiomics-based predictions of response to neoadjuvant chemotherapy - how are they affected by variations in tumour delineation?

Sepideh Hatamikia [1, 2]*, Geevarghese George [1], Florian Schwarzhans [1], Amirreza Mahbod [1], Ramona Woitek [1]

[1] Research Center for Medical Image Analysis and Artificial Intelligence (MIAAI), Department of Medicine, Danube Private University, Krems, Austria, Rathausplatz 1
AT-3500 Krems-Stein

[2] Austrian Center for Medical Innovation and Technology (ACMIT), Wiener Neustadt, Austria, Viktor Kaplan-Straße 2/1, 2700 Wiener Neustadt

* Corresponding author: Sepideh Hatamikia, MIAAI Research Center, Department of Medicine, Danube Private University, Krems, Austria, Rathausplatz 1
AT-3500 Krems-Stein, email: sepideh.hatamikia@dp-uni.ac.at, Telephone: +4366565293677.

**Abstract**

Manual delineation of volumes of interest (VOIs) by experts is considered the gold-standard method in radiomics analysis. However, it suffers from inter- and intra-operator variability. A quantitative assessment of the impact of variations in these delineations on the performance of the radiomics predictors is required to develop robust radiomics based prediction models. In this study, we developed radiomics models for the prediction of pathological complete response to neoadjuvant chemotherapy in patients with two different breast cancer subtypes based on contrast-enhanced magnetic resonance imaging acquired prior to treatment (baseline MRI scans). Different mathematical operations such as erosion, smoothing, dilation, randomization, and ellipse fitting were applied to the original VOIs delineated by experts to simulate variations of segmentation masks. The effects of such VOI modifications on various steps of the radiomics workflow, including feature extraction, feature selection, and prediction performance, were evaluated. Using manual tumor VOIs and radiomics features extracted from baseline MRI scans, an AUC of up to 0.96 and 0.89 was achieved for human epidermal growth receptor 2 positive and triple-negative breast cancer, respectively. For smoothing and erosion, VOIs yielded the highest number of robust features and the best prediction performance, while ellipse fitting and dilation lead to the lowest robustness and prediction performance for both breast cancer subtypes. At most 28% of the selected features were similar to manual VOIs when different VOI delineation data were used. Differences in VOI delineation affects different steps of radiomics analysis, and their quantification is therefore important for development of standardized radiomics research.

1. Introduction

Breast cancer is one of the most frequent cancer types in women, with an increasing incidence rate worldwide [1]. Histopathological examination of breast biopsies is the gold standard for cancer diagnosis and grading, but has limited capabilities for capturing tumour heterogeneity [2–4]. Non-invasive imaging techniques such as magnetic resonance imaging (MRI) can capture the intra- and inter-tumour heterogeneity and are widely used for breast cancer screening, diagnosis, and local staging [4,5]. Based on the cancer subtype and stage, different treatment plans such as neoadjuvant chemotherapy (NAC) are recommended to patients [6]. With the advent of computerized approaches such as advanced machine learning (ML)-based techniques, tumour characteristics can be extracted from images in an automatic manner in addition to the clinical interpretation of MRI scans by specialised radiologists that is prone to inter- and intraobserver variability [3,5,7,8]. Recently developed deep learning approaches can even directly classify or categorize MR images without handcrafted feature sets [7,9]. However, most of these approaches are data hungry, and the produced results are not interpretable, which makes it difficult to utilize them in clinical setting [10,11]. Radiomics refers to sets of computational and interpretable image features that can be used for the development of machine-learning based predictors of clinical outcome and could thus improve clinical decision-making. As shown in Figure. 1, the generic workflow of a radiomics-based analysis pipeline includes manual or automatic tumour segmentation [5,12], radiomic feature extraction from delineated volumes of interest (VOI) [5,13], feature selection to reduce the feature number by removing redundant or unimportant features [14,15], and a machine learning model [16,17] to classify images based on the clinical question (e.g., prediction of pathological complete response (PCR) to NAC) [11,18]. Although radiomics models have shown excellent results in characterizing breast lesions and predicting the outcome in breast cancer patients, the robustness of such models is highly dependent on the delineation (=segmentation) of lesions as segmentation masks are typically used to extract quantitative radiomics features. While automatic methods can be used for lesion segmentation of a radiomics model with acceptable predictive performance [19,20], manual delineation by experts is still considered the gold standard. However, this method is subjective and suffers from inter- and intra-operator variability. Therefore, quantitative assessment of variations in delineation and their impact on the radiomics model performance is required. The effects of manipulating segmentation masks on the robustness and the classification performance of ML-based predictive models have been investigated to some extent for different imaging modalities. Zhang et al. [21] applied different morphological operations (erosion, dilation, and smoothing) on the segmentation masks of MR images for two disease groups (metastasis in nasopharyngeal carcinoma and sentinel lymph node metastasis) and investigated their impacts on a predictive radiomic model. They showed that extensive changes to VOIs led to fewer reproducible features. VOI modification of only 3mm had no significant effect on the AUC classification scores for most cases, while wider dilation (5mm or 7mm) affected predictive performance differently in different diseases. Mahbod et al. [22] showed in dermatoscopic skin lesion images that cropping images (utilizing dilated lesion segmentation masks) improved classification compared to original images. Kocak et al. [23] investigated the influence of a morphological erosion operation (2 mm) on segmentation masks for computed tomography. The erosion yielded better reproducibility of textural features than the original segmentations but led to poorer classification performance. Lu et al. [24] tried two manual delineation methods (minimum and maximum) on MR images of patients with rectal cancer. The study showed good robustness for most of the extracted radiomic features for both approaches, with significantly better prediction performance of the maximum method compared to the minimum method.

In this study, first, we developed radiomic models to predict PCR to NAC in patients with human epidermal growth factor 2 (HER2) positive or triple-negative breast cancer (TNBC). A two-step feature selection method was proposed including, firstly, with Univariate Feature Selection (UFS) and correlation coefficients and, secondly, with different filter-, embedded- and wrapper- based feature selection algorithms. Combinations of different feature selection methods with different ML models were investigated for the prediction of PCR based on radiomics from pre-treatment contrast-enhanced MRI. Thereafter, we thoroughly investigated the effects of different VOI permutations on the feature robustness, feature selection and prediction performance of chemotherapy response using the developed radiomics models for the two breast cancer subtypes separately. Different mathematical operations (erosion, smoothing, dilation, randomization, and ellipse fitting) were applied to the original VOIs to generate diverse variations in the segmentation masks, and their effects on radiomics analysis were assessed. In addition, we quantified the differences in radiomics features due to VOI modifications as a reference for the radiomics analysis.

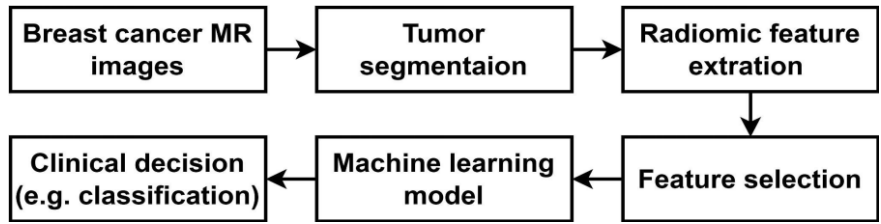

Figure 1: Generic workflow of radiomic-based image analysis.

## 2. Materials and methods
### 2.1. Dataset

We used the multicentre Investigation of Serial Studies to Predict Your Therapeutic Response with Imaging and moLecular Analysis (I-SPY1 TRIAL) breast MRI dataset. This is an open-access dataset that includes contrast-enhanced MRI and tissue-based biomarkers to predict PCR and relapse-free survival (RFS) [25–27]. MRI scans were performed at four different imaging time points (T1-T4): T1 = pre-treatment/baseline within four weeks prior to chemotherapy initiation, T2 = early treatment, day 1/cycle 2, T3 = between regimens, and T4 = prior to surgery). MRI scans were performed using 1.5-tesla field strength scanners (with dedicated breast radiofrequency coils). The imaging protocols included sagittal dynamic contrast-enhanced T1-weighted gradient echo sequences with TE = 4.5 ms, TR≤20 ms, 16-18 cm field of view, flip angle ≤ 45º, minimum matrix 256x192, slice thickness ≤ 2.5 mm, and 64 slices were used in this study to extract radiomics features. For each patient in the dataset a binary pCR score and three-level receptor status were available (see Table 1) [27]. A pCR score of 1 corresponds to complete response (no invasive cancer left on the surgical specimen obtained during surgery) and 0 otherwise. The hormone receptor positive and HER2 negative (HR+/HER2-) cohort was excluded from this study due to the low fraction of patients that showed complete response (Table 1).

| 3-level Hormone Receptor (HR) Status | Patients with pCR score | pCR=1 fraction |
|---|---|---|
| HR positive and HER2 negative (HR+/HER2-) | 65 | 0.11 |
| HER2 positive (HR-/HER2+) | 49 | 0.47 |
| Triple negative (TNBC) | 37 | 0.35 |
| Total | 151 | 0.31 |

Table 1: Summary of the number of patients, the prevalence (pCR=1) in the dataset separated by the Hormone Receptor (HR) Status.

For radiomics analyses, radiomics features were extracted from the VOI based on the segmentation masks and the first phase of the dynamic contrast-enhanced T1-weighted MRI. From the segmentation masks provided for the I-SPY1 dataset [4,25–28], we used the manually annotated structural tumour volume (STV) to extract radiomic features (Section 2.2) [28]. Several preprocessing steps, including bias field correction (N4 bias field correction with Simple ITK toolkit) [28,29], resampling to 1 mm$^3$ isotropic resolution, and z-score normalization were applied to the MRI data before radiomics feature extraction.

### 2.2. Radiomics feature extraction

Radiomics features can be divided into several categories, including shape, first-order, and higher-order features. Shape features describe the geometric properties of the lesion, such as its size, volume, and compactness. First-order features describe statistical properties of the voxel intensities within the lesion, such as mean, variance, skewness, and kurtosis. Higher-order features describe the spatial relationships between groups of voxels within lesions, (i.e. texture) which can be quantified using various methods such as gray-level co-occurrence matrices (GLCM), gray-level run-length matrices (GLRLM), gray-level size zone matrix (GLSZM), neighbouring gray tone difference matrix (NGTDM) and gray-level dependence matrix (GLDM) [30,31]. In this study, we used the Pyradiomics toolkit (version 3.0.1) [32] to extract radiomics features from the MRI. Pyradiomics is an open-source software package written in Python that provides a wide range of standard feature extraction methods for medical images. Pyradiomics settings selected for feature extraction were set to include all feature classes - adding up to a total of 107 features. NGTDM features were excluded as they were not predictive in our preliminary experiments, resulting in 102 features. Image type was set to original without any further preprocessing in the extraction step and the bin count was fixed to 100 bins. The feature extraction method was set to be separated – processing every lesion in an image separately – and only the radiomics features of the largest lesion per subject were used for further analysis.

### 2.3. Feature selection framework

Feature selection is a common step in many radiomics based workflows [33–40]. Given the high dimensionality of the input data in radiomics, $X \in R^{m \times n}$, with the number of features $n \approx 100 \gg m$, the number of patients, feature selection methods are employed to remove redundant and irrelevant features, to improve classification performance [40–42]. Feature selection methods are generally categorized as filter, wrapper and embedded algorithms [43]. Filter based algorithms rely on the general statistics of the features themselves without involving any learning algorithm [38,43]. It is argued that these features, which are selected based on their general characteristics, may not perform well in the final training because a learning algorithm is not involved in the feature selection phase [43]. To address this, wrapper methods have been introduced where we evaluate the predictive performance of a learning algorithm based on a subset of features [42,43]. Such methods are search-based algorithms and therefore the computational complexity to evaluate the performance of all possible subsets of features is very high [41,43]. Embedded methods try to overcome these drawbacks by making use of the inherent learning process to find the best features. For example, Least Absolute Shrinkage and Selection Operator (LASSO) is often used as a regularized sparse classification algorithm, which by definition encourages the coefficients of the model to be zero [44,45]. The remaining features with non-zero coefficients are deemed important for prediction; this way LASSO classifier masquerades as an embedded feature selection algorithm [41,44,45].

Figure 2 represents our proposed two-stage feature selection workflow. Radiomics features extracted from the breast MRI scans are processed in three steps: (A) the largest feature set, which is a set including all features $f_A$, (B) a reduced set of features $f_B$ which is found using Univariate Feature Selection (UFS) (Stage I), (C) and finally the best feature set $f_C$ selected by the different feature selection algorithms (Stage II). Features selected at each step are a subset of the previous feature set, i.e $f_C \in f_B \in f_A$ and $f_A$ and $f_C$ are used for training and validation to report the prediction performance (see Section 2.4).

In Stage I, in order to eliminate highly correlated, redundant features and to limit the search space, we first calculate the absolute spearman correlation matrix for all features inside a 5-fold cross validation (CV) loop [45]. Within each fold, features with a correlation coefficient $r$, below a given cutoff $r_c$, are preserved under the assumption that they are relevant. Instead of simply removing the remaining correlated features ($r > r_c$), they are given a second chance based on their relevance for classification (see Algorithm 1). The feature relevance is decided by UFS, which naively looks for the association or influence of a single independent variable (feature) with the dependent variable (target) using a classifier. As each feature is taken from a set of correlated features, the feature here is clearly not independent, but our goal is only to test for relevance of a feature that may be lost due to a harsh thresholding. A union of features selected from all folds is taken to form the set $f_B$, which is fed into the main feature selection algorithms (Stage II). Note that $r_c$ is a hyperparameter that can be further optimized. In this study we have used 9 feature selection methods including F-Score, Relief, Mutual Information (MI), Gini Importance, LASSO, Genetic Algorithm (GA), Sequential Backward Search (SBS), Sequential Forward Search (SFS), and Recursive Feature Elimination (RFE) [42,46]. The selected feature set from each of the above algorithms is evaluated for classification performance (Area Under the ROC Curve (AUC)) and the algorithm that maximizes this score is selected as the best feature selection algorithm and provides the best feature set $f_C$.

**Algorithm 1: Stage I feature selection with UFS**

input: **Set of all features from radiomics feature extraction**
output: **Reduced set of features that can be further analyzed,** $f$

1. $f_s \leftarrow$ Set of correlated features with Spearman-$r > r_c$;
2. N $\leftarrow$ Number of items in $f_s$;
3. **for** i $\leftarrow 1\ to\ N$ **do**
       $f_{corr} \leftarrow$ correlated features with $f_s[i]$ with Spearman-$r > r_c$;
       s $\leftarrow$ UFS score of $f_s[i]$;
       $s_{corr} \leftarrow$ UFS scores of correlated features $f_{corr}$;
       **if** s $\geq \max(s_{corr})$ **then**
         f $\leftarrow$ f $\cup$ $f_s[i]$;
       **end**
   **end**
4. $f_o \leftarrow$ Set of uncorrelated features with Spearman-$r \leq r_c$;
5. $f \leftarrow f \cup f_o$;

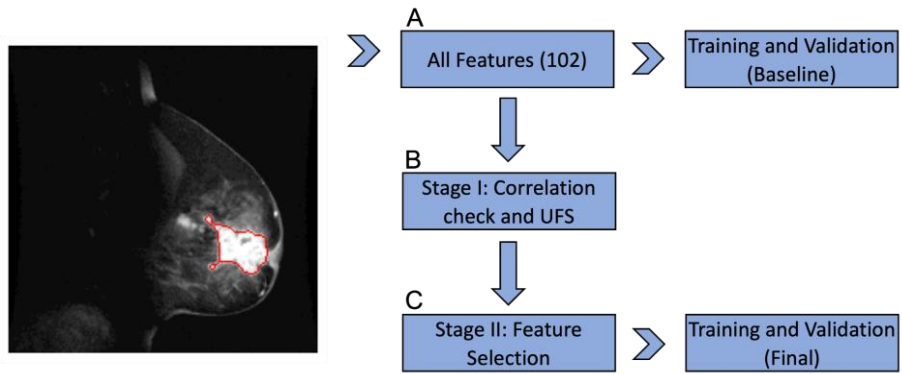

Figure 2: Breast MRI with manual segmentation of the largest lesion of multifocal breast cancer (left panel) and diagram showing processing of extracted radiomics features in three steps (middle and right panels): (A) all features are kept, (B) a reduced set of features is found, (C) and finally the optimal number of feature set is found. The features found in A and B are used for training and validation.

### 2.4. Training and prediction of response

The dataset, for a given breast cancer subtype (see Table 1), was split into an 80/20 ratio in a stratified manner, ensuring proportionate class distributions. The first split was used for feature selection (see Section 2.3) and training. The second split was kept aside as a hidden/test set for final evaluation. Evaluation was performed using AUC score, sensitivity, and specificity. A stratified k-fold cross CV loop with $k = 5$ was used for training. During training, each feature in the training fold was centered and scaled by calculating the relevant statistics (mean and standard deviation). The extracted statistics were used to further transform the validation fold. The PCR labels were already binarized by the data provider with 0 (non-complete response) and 1 (complete response), therefore target encoding was not necessary. Each training fold generates a fitted model, and an average CV score is reported along with its standard deviation. The fitted model and statistics from each of the $k$-training folds were saved for testing. The final evaluation was performed on the test set using the ensemble of $k$ models after scaling with the saved training statistics. The average score and the standard deviation are reported. Due to the limited number of samples and high dimensionality, in our experiments we utilized two linear ML models (classifiers) [45]: 1) Logistic Regression (LR) with L1 and L2 regularization, commonly known as ElasticNet, with balanced class weights and the SAGA solver [44,47] 2) Shrinkage enabled Linear Discriminant Analysis (LDA) with the eigenvalue decomposition solver [45]. The preprocessing and modelling pipelines were developed using the scikit-learn (version 1.0.2) [46,48]. Important hyperparameters for LR that control the regularization such as `l1_ratio` and C, and for LDA such as shrinkage were tuned within the CV loop after UFS [45]. Hyperparameter tuning was performed using the Optuna (version 3.0.0) [49].

### 2.5. Modifications of regions of interest

In order to investigate the effect of varying delineations of the VOIs on different radiomics features and predictive models, we have modified the segmentation contours in different ways and analysed the robustness of the features and models with respect to the change in contour. The VOI modifications used were as follows:

Systematic and uniform enlargement/shrinking of the lesions using morphological operators, to assess the effect of over- or underestimating the outline of the region; smoothing of the lesion contour to simulate a less-detailed segmentation approach by getting rid of sharp corners and small details in the outline; randomization of the contour outline by modifying the outline randomly by a small value to simulate random segmentation differences that may occur between multiple readers performing manual segmentations. Additionally, we have applied an approximation of an ellipsoid to investigate whether some radiomics features stay robust even if the segmentation is only a simple ellipsoid roughly encompassing the lesion. For all operations the volumes were resampled to an isotropic spacing in all 3 dimensions (choosing the lowest pixel spacing among the 3 axes for all 3 axes) using nearest neighbour interpolation and after processing, the modified volumes were resampled back to their original spacing. The isotropic resampling was applied for the operations to have the same magnitude in all 3 axes.

**(a) morphological operators**

Morphology is a mathematical concept which is based on analysis of the shape of objects. It uses an input image A and a binary structural element B.

Morphological operators act on the input image and change its pixel values based on the method defined and considering the shape of the given kernel. Similar to a sliding window average operation the kernel is shifted across all pixels in the given image. For each position (X, Y, Z) the kernel passes the corresponding morphological operation and is applied over all pixels covered by the kernel. One important aspect to morphological operations is the shape and size of the chosen kernel, as this directly affects the shape and size of the enhanced/reduced features. For uniform manipulation a circular or spherical kernel shape is usually chosen, but the kernel can in theory have any arbitrary shape. Mathematically the two main morphological operations (Figure. 3) are defined as follows:

Morphological dilation of an image A by a structural element B is denoted as $A \oplus B$ and is defined as

$$A \oplus B = \{B_z \cap A \neq 0\}$$

This means that dilation is the amount of pixel positions z where the structural element B when shifted by an amount of z still overlays one or more pixels in A with a value of 1.

Morphological erosion of an image A by a structural element B is denoted as $A \ominus B$ and is defined as

$$A \ominus B = \{B_z \subseteq A\}$$

This means that erosion is the amount of pixel positions z where the structural element B when shifted by an amount of z overlays only pixels in A with a value of 1.

To perform the morphological operations on the VOI, spherical binary kernels with radiuses of 1 mm and 2 mm were chosen (denoted as dilation 1, dilation 2, erosion 1, erosion 2), resulting in two different sizes for the eroded and dilated VOI modified images.

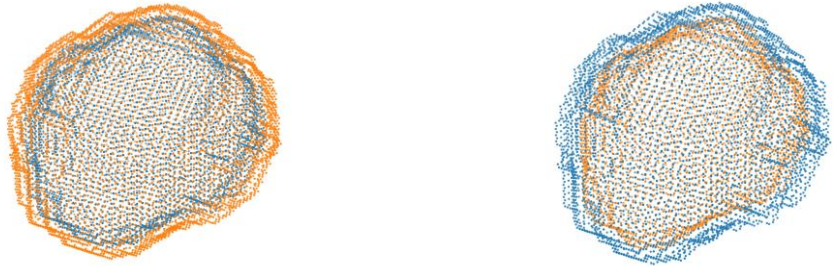

Figure 3: Two-dimensional point cloud representation of an example lesion. Blue dots represent original lesion VOI, orange dots represent VOI after morphological dilation was applied causing the VOI to grow larger (left) and after morphological erosion was applied causing the VOI to shrink (right).

**(b) smoothing**

Smoothing operations on the binary VOI were performed via gaussian smoothing, implemented via a sliding window function based on the equation

$$G(X,Y,Z) = \frac{1}{\sqrt{2\pi\sigma^2}} e^{-\frac{x^2+y^2+z^2}{2\sigma^2}}$$

With (x/y/z) being the distance in the 3 respective axes to the current center point (X,Y,Z) and $\sigma$ denoting the magnitude of smoothing. Finally the new segmentation $S_n$ is created as

$$S_n(X,Y,Z) = \{1 \ \ if \ G(x,y) > 0.5; \ 0 \ otherwise$$

Smoothing (Figure 4) was performed on the VOI using two different magnitudes with σ values of 1 mm and 2 mm respectively, resulting in two versions of the smoothed binary image (denoted as smoothing 1 and smoothing 2).

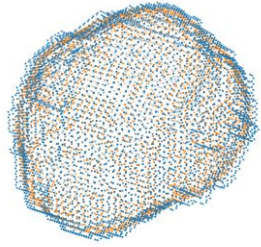

Figure 4: Two-dimensional point cloud representation of unmodified VOI (blue dots) and smoothed VOI (orange dots). Sharp corners (i.e. upper right corner) have been smoothed out by the smoothing operator.

**(c) Randomization**

We propose a randomization approach which involves modifying the outline of the VOI by moving each point inward or outward of the VOI by a random, yet smooth, distance (Figure. 5). This resulted in a new outline that followed the general shape of the original VOI but introduced some variability. To achieve this, we first converted the outline of the VOI into a mesh object using the marching cubes algorithm implemented in the scikit-image Python package (version 0.20.0) [46]. We then used the open3d Python package (version 0.16.0) [50] to manipulate the mesh and compute the normal vectors of all vertices. Specifically, we smoothed the mesh to obtain a more accurate representation of the rough shape of the VOI and computed the normal vectors of each vertex based on this smoothed copy. To modify the outline, we used Perlin Noise. Perlin Noise is a type of gradient noise that generates smooth and natural-looking patterns. It works by generating a series of pseudo-random gradients at different points in space and interpolating between them. The resulting noise values have a smooth transition between neighbouring points, which helps avoid sharp spikes or artifacts in the final modification. The magnitude and direction of the shift for each vertex were determined by calculating the Perlin Noise value of the normalized coordinates of the vertex and multiplying it by the maximum distance that the modification should incorporate. By normalizing the coordinates, we ensured that the noise function was scale-invariant and could be used on VOIs of different sizes. The maximum distance parameter controlled the overall magnitude of the modification and allowed us to tune the level of variability introduced.

We simulated different magnitudes of alteration by setting the maximum distance parameter for the modification to values of 1 mm and 2 mm, resulting in two different versions of the randomized VOI outline per image (randomization 1, randomization 2).

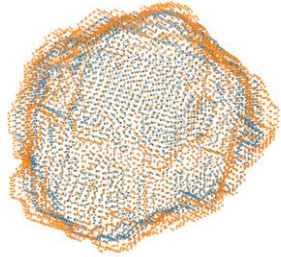

Figure 5: Two-dimensional point cloud representation of unmodified VOI (blue dots) and randomized VOI (orange dots). Over the entire surface of the lesion patches can be seen where the modified VOI is bulging out from the original VOI, and also patches where it bulges in. Thus, the overall rough shape of the VOI remains similar but with variations in the exact boundary.

### (d) Ellipsoid fitting

While mathematically optimal ellipsoid fitting can be a complex and computationally intensive task, our goal is to simulate the process of an operator drawing an approximate ellipsoid around the lesion (Figure. 6). To achieve this, we first convert the outline of the VOI into a mesh object using the marching cubes algorithm and then smooth it using the open3d package to obtain a more accurate representation of the rough shape of the VOI. In practice, a grader would typically estimate the major axis of the ellipsoid by approximating the longest part of the VOI, and then rotate and scale the perpendicular axes to best fit the shape of the VOI. To simulate this process, our implementation calculates the Euclidean distance between all pairs of vertices in the mesh and selects the pair with the highest distance as the start and end coordinates of the major axis. The two perpendicular axes are then rotated in 1° increments around the major axis, and the fit of the ellipsoid around each rotation is measured by the sum of squared differences between the ellipsoid and the VOI outline. The rotation and scale with the lowest error is selected as the optimal fit. Finally, a new mesh object is created using the approximated ellipsoid parameters and converted into a solid VOI which can be used for radiomics feature extraction.

Demonstration of the effect of different VOI manipulation on a 2D slice of the volume is shown in Figure. 7.

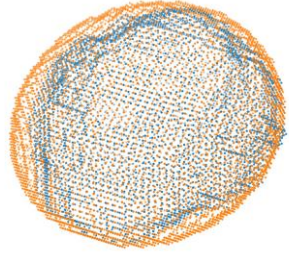

Figure 6: Two-dimensional point cloud representation of the unmodified VOI of the lesion (blue dots) and the approximated ellipsoid that best represents the shape and size of the VOI (orange dots).

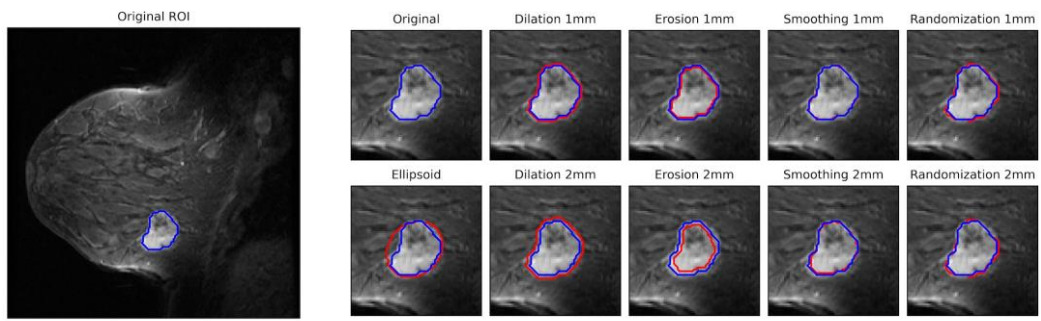

Figure 7: Demonstration of the effect of VOI manipulation on a single slice of the breast MRI. Blue contour shows original volume of interest (VOI) outline and red contour shows the corresponding operation.

## 2.6. Radiomics feature robustness

In order to quantify the robustness of individual radiomics features against changes in the segmentation of the VOI we calculated the Intraclass Correlation Coefficient (ICC). The ICC assesses the reliability of ratings by comparing the variability of different ratings of the same subject to the total variation across all ratings and subjects. VOIs obtained via the aforementioned VOI manipulation techniques can be interpreted as VOIs obtained from a different reader who, for example for morphological dilation, systematically delineates lesions larger than the original reader. Radiomics features obtained from a modified VOI can be compared with radiomics features from the unmodified VOI using the ICC as a measure of reliability or robustness. This comparison is performed for the values of every extracted radiomics feature for all subjects for each VOI manipulation method separately against the corresponding extracted features of the unmodified VOI, providing one ICC value per VOI manipulation operator, per feature. For analysis, these ICC values are grouped into the different radiomics feature categories (shape, first order and texture) to better understand which types of radiomics features are robust against which type of VOI manipulation. The proportion (in percentage) of each feature category with ICC above 0.9 is reported.

## 2.7. Evaluation of performance of the predictive models using different modifications of volumes of interest

In a first analysis, we evaluated the performance of the achieved predictive models using the original manual segmentations (Section 3.1) also for data with delineation using different VOIs (described in Section 2.5). In this case, the model (including features and classifier) as selected based on manual segmentations was used, but the radiomics features were extracted from modified VOIs for the test dataset only. This analysis provides insight into the robustness of the developed radiomics predictive models (Section 3.3) over different modifications of the VOIs in test data. In a second analysis, we let the model select the best performing features based on the radiomics features extracted from each VOI modification. So, for each modification of the VOIs, a different model including different classifiers and features could be selected (Section 3.3). With this approach, we were able to evaluate whether the features selected from data using different VOIs show differences or not.

A diagram showing the general proposed workflow is presented in Figure. 8.

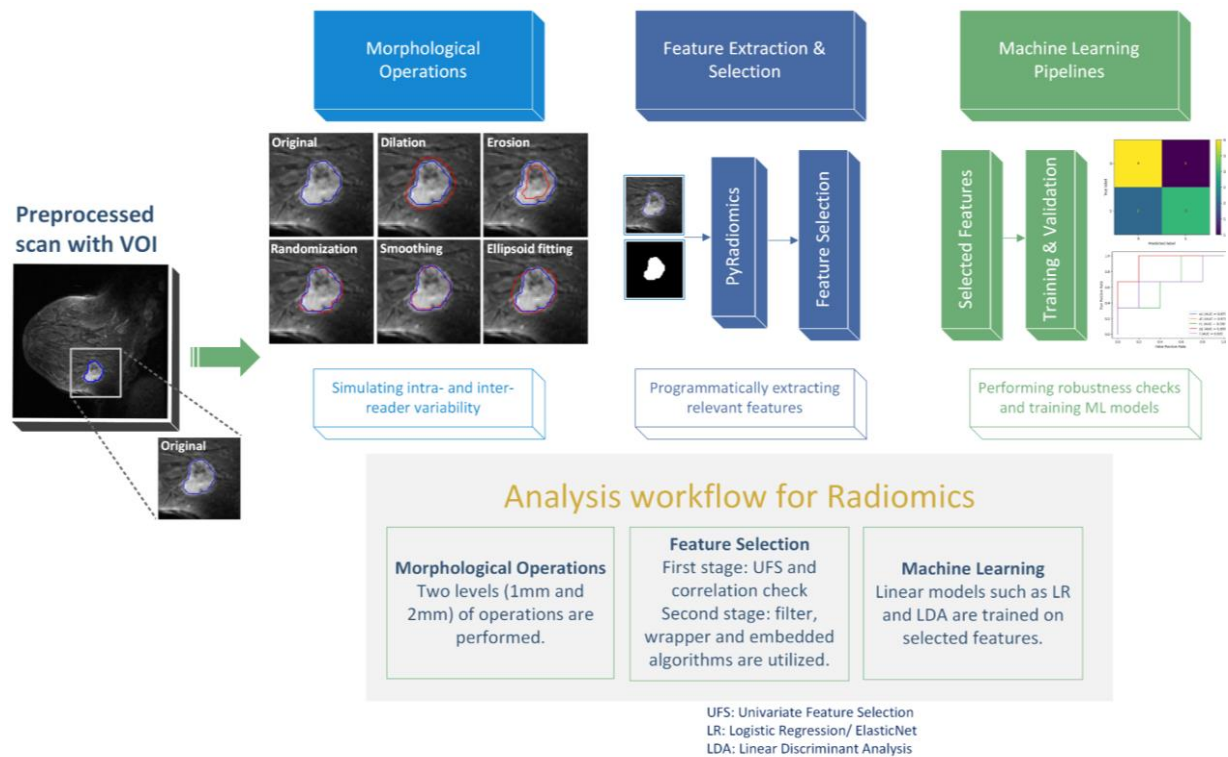

Figure 8: A diagram presenting the general proposed radiomics workflow.

## 3. Results
### 3.1. Feature selection and prediction performance analysis

The AUC, sensitivity, and specificity of training and test datasets related to the three best performing radiomics models as well as selected features for the two subtypes of breast cancer, HER2+ and TNBC, are shown in Tables 1 and 2, respectively. An AUC of up to 0.96 and 0.89 was achieved for test datasets for HER2+ and TNBC, respectively. In addition, the FS algorithms LASSO, SFS, and RFE were selected for both subtypes as the best FS methods. The number of selected features was up to 7.

Table 2. The Area under the curve (AUC), sensitivity (SE), specificity (SP), and selected features using the three best radiomics models selected for HER2+ breast cancer.

| Three best FS Algorithms | Classifier | Train dataset AUC, SE, SP | Test dataset AUC, SE, SP | Number of features | Features |
|---|---|---|---|---|---|
| **LASSO** | LDA | 0.82; 0.72; 0.80 | 0.96; 0.68; 1.00 | 7 | ['firstorder_InterquartileRange', 'firstorder_Variance', 'glrlm_ShortRunHighGrayLevelEmphasis', 'glcm_DifferenceVariance', 'glszm_LargeAreaHighGrayLevelEmphasis', 'glcm_MCC', 'shape_SurfaceVolumeRatio'] |
| **SFS** | LR | 0.62; 0.43; 0.80 | 0.70; 0.44; 0.96 | 1 | ['firstorder_Uniformity'] |
| **RFE** | LR | 0.59; 0.63; 0.54 | 0.68; 0.68; 0.68 | 6 | ['firstorder_InterquartileRange', 'firstorder_TotalEnergy', 'gldm_SmallDependenceLowGrayLevelEmphasis', 'glszm_LargeAreaHighGrayLevelEmphasis', 'glszm_HighGrayLevelZoneEmphasis', 'shape_Sphericity'] |

Table 2. The Area under the curve (AUC), sensitivity (SE), specificity (SP), and selected features using the three best radiomics model selected for TNBC.

| Three best FS Algorithms | Classifier | Train dataset AUC, SE, SP | Test dataset AUC, SE, SP | Number of features | Features |
|---|---|---|---|---|---|
| SFS | LDA | 0.94; 0.80; 0.95 | 0.89; 0.60; 0.80 | 7 | ['firstorder_10Percentile', 'firstorder_Skewness', 'glcm_ClusterShade', 'glrlm_ShortRunLowGrayLevelEmphasis', 'glcm_MCC', 'glrlm_LongRunLowGrayLevelEmphasis', 'shape_Elongation'] |
| LASSO | LDA | 0.93; 0.70; 0.95 | 0.85; 0.60; 0.80 | 5 | ['firstorder_Minimum', 'firstorder_Skewness', 'glcm_ClusterShade', 'glrlm_ShortRunLowGrayLevelEmphasis', 'gldm_LowGrayLevelEmphasis'] |
| RFE | LDA | 0.94; 0.80; 0.95 | 0.85; 0.47; 0.80 | 9 | ['firstorder_Minimum', 'firstorder_10Percentile', 'firstorder_Skewness', 'glcm_ClusterShade', 'glcm_MCC', 'glrlm_LowGrayLevelRunEmphasis', 'gldm_LowGrayLevelEmphasis', 'glcm_DifferenceVariance', 'glrlm_LongRunLowGrayLevelEmphasis'] |

### 3.2. Feature robustness analysis

The ICC values for all VOI variants calculated for each feature individually are shown in Figures 9 and 10. In addition, the percentage of features with ICC values above 0.9 for both cancer subtypes are calculated for all features and for different feature groups (shape, first order, texture) (Tables 3 and 4). The ICC values were highest for smoothing 1, dilation 1, and randomization 1 for both cancer subtypes, with smoothing 1 showing the highest robustness. With these modifications, all shape features (100%) had ICC values above 0.9. In the HER2+ group, 97% of texture features but only 83% of first-order features had ICC values above 0.9. A similar pattern was observed for TNBC: (a higher proportion of texture features (92%) showed an ICC above 0.9 compared to first order features (89%)). Ellipsoid fitting, erosion 2, and dilation 2 showed the largest effects on feature values, with ellipse fitting showing the lowest robustness (11% and 13% of all features with an ICC above 0.9 for HER2+ and TNBC, respectively). For these three operators, shape features showed better robustness than first order features and texture features. Overall, smoothing 1 resulted in the highest number of robust features, while ellipse fitting had the lowest ICC range for both cancer subtypes.

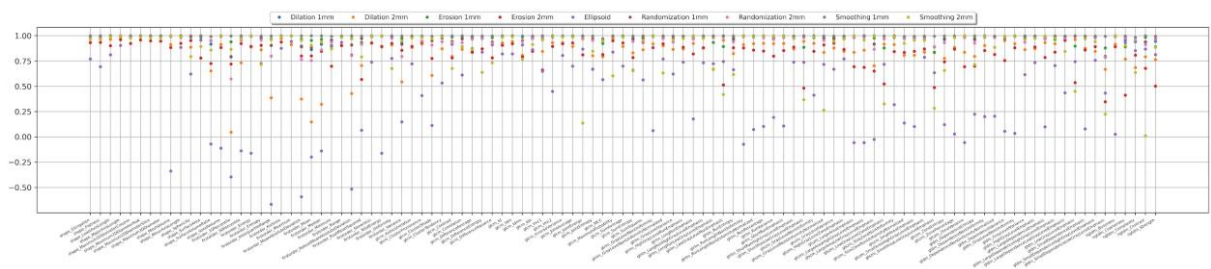

Figure 9: ICC values related to each radiomics feature for all VOI modifications for HER2+ breast cancer.

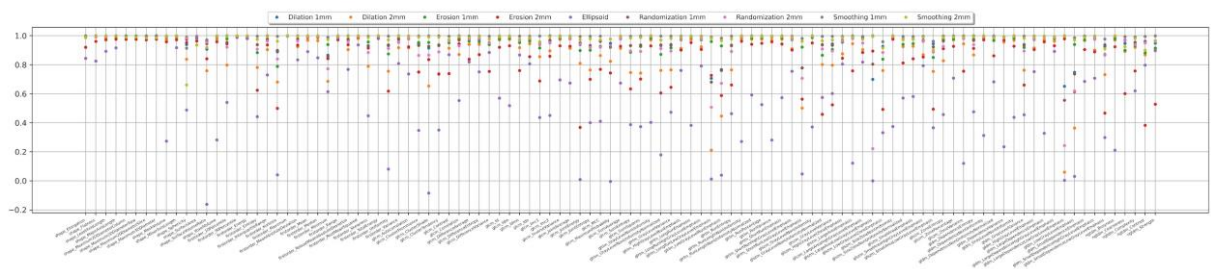

Figure 10: ICC values related to each radiomics feature for all VOI modifications for TNBC.

Table 3. The percentage of radiomics features (all, shape, first order and texture) with ICC above 0.9 for HER2+ breast cancer.

| Features | Dilation 1 | Dilation 2 | Erosion 1 | Erosion 2 | Ellipsoid | Randomization 1 | Randomization 2 | Smoothing 1 | Smoothing 2 |
|---|---|---|---|---|---|---|---|---|---|
| **All** | 95% | 44% | 93% | 30% | 11% | 95% | 91% | 100% | 67% |
| **Shape** | 100% | 86% | 100% | 71% | 43% | 100% | 100% | 100% | 79% |
| **First order** | 83% | 33% | 100% | 39% | 11% | 83% | 61% | 100% | 72% |
| **Texture** | 97% | 39% | 89% | 20% | 5% | 97% | 96% | 100% | 64% |

Table 4. The percentage of radiomics features (all, shape, first order and texture) with ICC above 0.9 for TNBC.

| Features | Dilation 1 | Dilation 2 | Erosion 1 | Erosion 2 | Ellipsoid | Randomization 1 | Randomization 2 | Smoothing 1 | Smoothing 2 |
|---|---|---|---|---|---|---|---|---|---|
| **All** | 93% | 61% | 88% | 54% | 13% | 93% | 80% | 100% | 98% |
| **Shape** | 100% | 86% | 100% | 93% | 50% | 100% | 100% | 100% | 93% |
| **First order** | 89% | 67% | 83% | 78% | 33% | 89% | 83% | 100% | 100% |
| **Texture** | 92% | 55% | 87% | 41% | 1% | 92% | 76% | 100% | 99% |

### 3.3. Model performance results using different modifications of volumes of interest

For the best performing radiomics models for both HER2+ and TNBC, we compared the prediction results when using radiomics features extracted from the original manual segmentations with those when extracting features from modified VOIs (Section 2.7). The FS algorithm, subset of radiomics features and the classifier were selected for training using original manual VOIs, but we assessed the performance of the model on test data using radiomics features extracted from VOIs that had undergone the different modifications described above (Tables 5, 6). To quantify the change in performance, measured by the AUC score, we define $\Delta_{AUC} = \frac{AUC_2 - AUC_1}{AUC_1} \times 100$, where $AUC_2$ is the score after a given VOI modification and $AUC_1$ is the score obtained with the original manual VOIs.

As shown in Tables 5 and 6, for both breast cancer subtypes, the VOI modifications erosion 1 and both smoothing 1 and 2 resulted in the lowest test $\Delta_{AUC}$ meaning these delineations affected the prediction results in the test dataset the least and performed most similar to the original manual VOIs compared to other modifications. In contrast, the highest difference in prediction results compared to the original manual VOIs was achieved using ellipse fitting and dilation 2 operators in both cancer subtypes.

Table 5. The Area under the curve (AUC), sensitivity (SE), specificity (SP), train and test change (%) and average ICC of features related to a selected model using original manual VOIs as well as different VOI modifications for HER2+ breast cancer group, d1: dilation 1, d2: dilation 2, e1: erosion 1, e2: erosion 2, l: ellipsoid fitting, r1:randomization 1, r2:randomization 2, s1:smoothing 1, s2:smoothing 2. NA: not applicable, Avg: average.

| FS algorithm and classifier | Modification | Train AUC, SE, SP | Test AUC, SE, SP | Train $\Delta_{AUC}$ | Test $\Delta_{AUC}$ | Avg. ICC of selected features |
|---|---|---|---|---|---|---|
| LASSO, LDA | none | 0.82; 0.72; 0.80 | 0.96; 0.68; 1.00 | NA | NA | NA |
| LASSO, LDA | d1 | 0.66; 0.55; 0.76 | 0.72; 0.60; 0.96 | -20.10 | -25.00 | 0.94 |
| LASSO, LDA | d2 | 0.19; 0.12; 0.53 | 0.46; 0.36; 0.68 | -77.25 | -52.50 | 0.69 |
| LASSO, LDA | e1 | 0.67; 0.62; 0.66 | 0.88; 0.76; 0.84 | -18.16 | -8.33 | 0.96 |
| LASSO, LDA | e2 | 0.61; 0.48; 0.56 | 0.60; 0.64; 0.60 | -25.68 | -37.50 | 0.79 |
| LASSO, LDA | l | 0.48; 0.20; 0.65 | 0.39; 0.32; 0.52 | -41.00 | -59.17 | 0.22 |
| LASSO, LDA | r1 | 0.78; 0.67; 0.81 | 0.74; 0.60; 1.00 | -4.66 | -23.33 | 0.95 |
| LASSO, LDA | r2 | 0.73; 0.62; 0.70 | 0.69; 0.60; 0.92 | -10.55 | -28.33 | 0.91 |
| LASSO, LDA | s1 | 0.78; 0.77; 0.75 | 0.93; 0.64; 0.88 | -4.66 | -3.33 | 0.99 |
| LASSO, LDA | s2 | 0.80; 0.67; 0.70 | 0.80; 0.60; 0.84 | -3.04 | -16.66 | 0.91 |

Table 6. The Area under the curve (AUC), sensitivity (SE), specificity (SP), train and test change (%) and average ICC of features related to a selected model as well as different VOI modifications for TNBC, d1: dilation 1, d2: dilation 2, e1: erosion 1, e2: erosion 2, l: ellipsoid fitting, r1:randomization 1, r2:randomization 2, s1:smoothing 1, s2:smoothing 2, NA: not applicable, Avg: average.

| FS algorithm and classifier | Modification | Train AUC, SE, SP | Test AUC, SE, SP | Train $\Delta_{AUC}$ | Test $\Delta_{AUC}$ | Avg. ICC |
|---|---|---|---|---|---|---|
| SFS, LDA | none | 0.94; 0.80; 0.95 | 0.89; 0.60; 0.80 | NA | NA | NA |
| SFS, LDA | d1 | 0.80; 0.60; 0.80 | 0.80; 0.40; 0.96 | -15.05 | -10.44 | 0.89 |
| SFS, LDA | d2 | 0.76; 0.60; 0.68 | 0.35; 0.33; 0.56 | -19.47 | -61.18 | 0.68 |
| SFS, LDA | e1 | 0.84; 0.60; 0.90 | 0.92; 0.73; 0.80 | -10.62 | 2.99 | 0.95 |
| SFS, LDA | e2 | 0.74; 0.50; 0.90 | 0.97; 0.80; 0.88 | -21.23 | 8.96 | 0.81 |
| SFS, LDA | l | 0.54; 0.30; 0.85 | 0.35; 0.33; 0.84 | -42.48 | -61.19 | 0.31 |
| SFS, LDA | r1 | 0.86; 0.60; 0.85 | 0.83; 0.40; 1.00 | -8.86 | -7.46 | 0.88 |
| SFS, LDA | r2 | 0.81; 0.50; 0.85 | 0.45; 0.33; 0.92 | -14.16 | -49.25 | 0.82 |
| SFS, LDA | s1 | 0.92; 0.70; 0.85 | 0.93; 0.67; 0.80 | -2.65 | 4.48 | 0.99 |
| SFS, LDA | s2 | 0.92; 0.80; 0.85 | 0.85; 0.73; 0.80 | -2.65 | -4.48 | 0.98 |

In a different approach, we let the model select the best features based on the radiomics features extracted from each VOI modification (Section 2.7). In this case, we observed that for different VOI modifications different feature selection algorithms performed best and that different features were selected (Tables 7 and 8), highlighting the strong effects of variations in VOI delineation on the feature selection process. We report the percentage of the common features (denoted as $f_c$ in Tables 7, 8) which is calculated by dividing the number of common features after applying each operation by the number of selected features using original manual VOIs. For both groups no more than 28% of features were the same when using different VOI modifications compared to original manual VOIs. The common features selected using original and modified VOIs are shown in Figures 11 and 12. We also reported the average ICC values over the selected features (Tables 5-8). We observed the lowest average ICC value (lowest robustness) for ellipsoid fitting and dilation 2, which matches the results showing the highest test-change achieved by these operators. In addition, the highest average ICC value for smoothing operators and erosion 1 is observed for both groups which also matches the lowest test-change achieved for these operators. However, a direct relationship between the average ICC values and test-change could not be found for other cases.

Table 7. The Area under the curve (AUC), sensitivity (SE), specificity (SP), train and test change (%), number of features, fraction of common features and average ICC of features related to selected models when using data from different VOIs modification for HER2+ breast cancer, d1: dilation 1, d2: dilation 2, e1: erosion 1, e2: erosion 2, l: ellipsoid fitting, r1:randomization 1, r2:randomization 2, s1:smoothing 1, s2:smoothing 2, NA: not applicable, Avg: average.

| Selected algorithm | Operation | Selected_ train_score SE, SP | Selected_ holdout_score SE, SP | Train $\Delta_{AUC}$ | Test $\Delta_{AUC}$ | Number of selected features using original masks | Number of common features after applying the operation | $f_c$ (%) | Avg. ICC |
|---|---|---|---|---|---|---|---|---|---|
| LASSO | none | 0.82; 0.72; 0.80 | 0.96; 0.68; 1.00 | NA | NA | 7 | NA | NA | NA |
| RFE | d1 | 0.93; 0.88; 0.80 | 0.76; 0.84; 0.56 | 12.70 | -20.83 | 21 | 2 | 9.52 | 0.97 |
| Gini | d2 | 0.72; 0.42; 0.73 | 0.53; 0.56; 0.32 | -12.28 | -45.00 | 2 | 1 | 50.00 | 0.39 |
| Gini | e1 | 0.62; 0.57; 0.67 | 0.81; 0.60; 0.76 | -24.67 | -15.83 | 7 | 2 | 28.57 | 0.96 |
| MI | e2 | 0.86; 0.65; 0.76 | 0.70; 0.44; 0.68 | 4.37 | -27.50 | 2 | 0 | 0 | 0.83 |
| Gini | l | 0.64; 0.42; 0.75 | 0.40; 0.16; 0.60 | -22.23 | -58.33 | 2 | 1 | 50.00 | 0.29 |
| Gini | r1 | 0.65; 0.33; 0.71 | 0.63; 0.44; 0.64 | -21.11 | -34.16 | 5 | 0 | 0 | 0.98 |
| RFE | r2 | 0.81; 0.67; 0.71 | 0.71; 0.44; 0.80 | -1.62 | -25.83 | 9 | 2 | 22.22 | 0.90 |
| LASSO | s1 | 0.74; 0.55; 0.77 | 0.78; 0.72; 0.80 | -9.64 | -19.17 | 8 | 4 | 50.00 | 0.99 |
| LASSO | s2 | 0.71; 0.58; 0.71 | 0.80; 0.60; 0.80 | -13.49 | -16.67 | 8 | 1 | 12.50 | 0.90 |

Table 8. The Area under the curve (AUC), sensitivity (SE), specificity (SP), train and test change (%), number of features, fraction of common features and average ICC of features related to selected models when using data from different VOIs modification for TNBC, d1: dilation 1, d2: dilation 2, e1: erosion 1, e2: erosion 2, l: ellipsoid fitting, r1:randomization 1, r2:randomization 2, s1:smoothing 1, s2:smoothing 2, NA:not applicable, Avg: average.

| Selected algorithm | Operation | Selected _train_score SE, SP | Selected _holdout_score SE, SP | Train $\Delta_{AUC}$ | Test $\Delta_{AUC}$ | Number of selected features using original masks | Number of common features after applying the operation | $f_c$ (%) | Avg. ICC |
|---|---|---|---|---|---|---|---|---|---|
| SFS | none | 0.94; 0.80; 0.95 | 0.89; 0.60; 0.80 | NA | NA | 7 | NA | NA | NA |
| RFE | d1 | 0.91; 0.70; 0.90 | 0.87; 0.40; 1.00 | -3.55 | -2.98 | 5 | 2 | 40.00 | 0.93 |
| LASSO | d2 | 0.86; 0.50; 0.80 | 0.61; 0.27; 0.84 | -8.86 | -31.34 | 4 | 2 | 50.00 | 0.77 |
| LASSO | e1 | 0.97; 0.60; 0.95 | 0.87; 0.53; 0.80 | 2.65 | -2.98 | 4 | 3 | 75.00 | 0.94 |
| GA | e2 | 0.94; 0.70; 0.95 | 0.79; 0.60; 0.80 | 0 | -11.93 | 5 | 2 | 40.00 | 0.87 |
| SBS | l | 0.87; 0.50; 0.95 | 0.63; 0.33; 0.88 | -7.96 | -29.84 | 3 | 0 | 0 | 0.74 |
| SFS | r1 | 0.88; 0.70; 0.90 | 0.67; 0.40; 1.00 | -6.20 | -25.36 | 15 | 6 | 40.00 | 0.92 |
| GA | r2 | 0.88; 0.70; 0.85 | 0.64; 0.33; 1.00 | -6.20 | -28.35 | 5 | 0 | 0 | 0.78 |
| GA | s1 | 0.97; 0.70; 0.95 | 0.89; 0.53; 0.92 | 3.54 | 0 | 5 | 1 | 20.00 | 0.99 |
| GA | s2 | 1.00; 0.80; 0.95 | 0.81; 0.40; 0.80 | 6.19 | -8.95 | 7 | 2 | 28.57 | 0.96 |

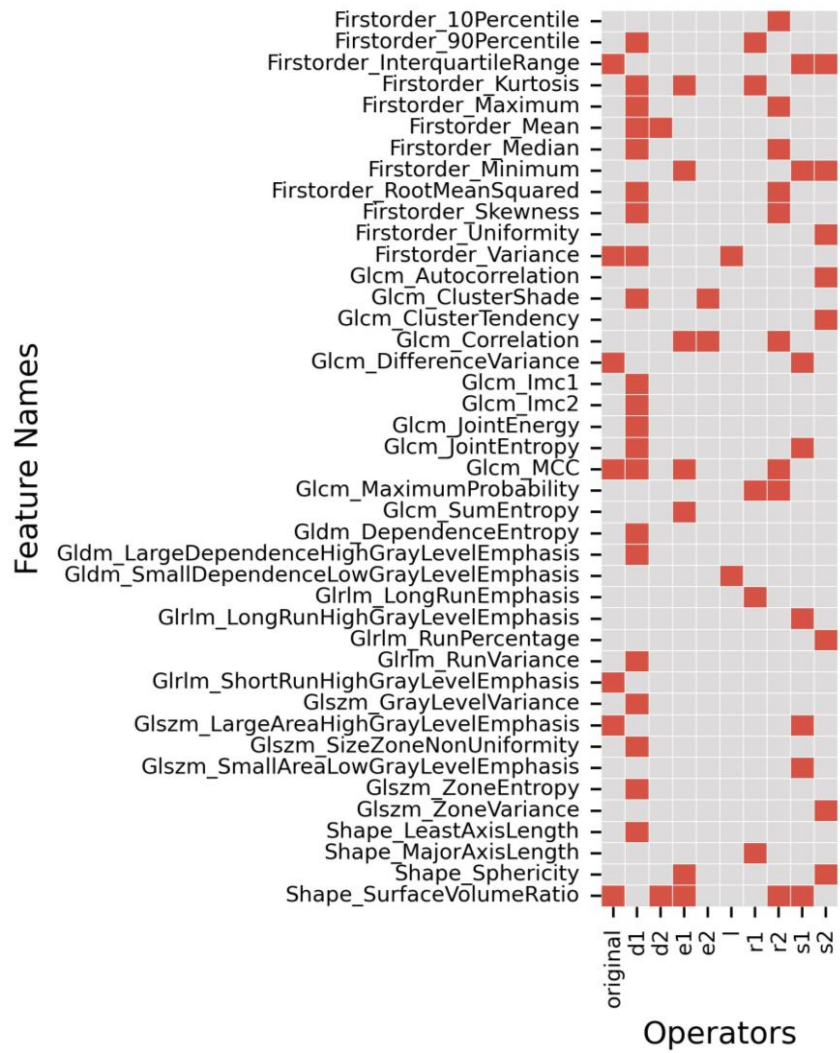

Figure 11: The common features selected by data using original manual and modified VOIs for HER2+ breast cancer.

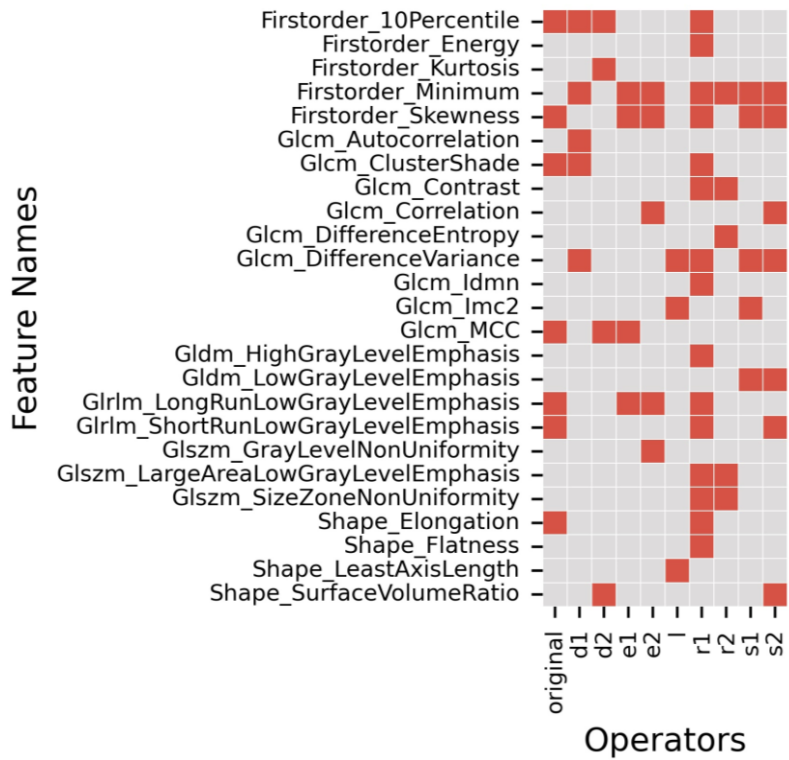

Figure 12: The common features selected by data using original manual and modified VOIs for TNBC.

## 4. Discussion

This is the first study, to comprehensively evaluate the effects of different tumor VOI modifications on radiomics analyses in HER2+ breast cancer and TNBC. We investigated how the introduction of systematic and randomised modifications of tumor delineation affect radiomics and radiomics-based predictors to provide references for radiomics studies in breast cancer and to aid the development of standardized radiomics research.

We developed radiomics-based prediction models of pCR for HER2+ breast cancer and TNBC based on pre-treatment breast MRI. In contrast to many other radiomics studies, we evaluated different feature selection methods and classifiers and achieved an AUC of 0.96 for HER2+ breast cancer and 0.89 for TNBC on held out test sets. These results are comparable to previously published work [51-53]. Our models, however, are based on radiomics of one MRI sequence only and reach AUC values comparable to prediction models combining clinical data and radiomics [51] or radiomics from multiple MRI sequences [52]. We identified LASSO and SFS as the best feature selection methods and LDA as the best classifier for the original manual VOIs of the ISPY1 data set.

We performed standard and more advanced modifications of those VOIs that are comparable to VOIs outlined by different radiologists, such as randomization, smoothing, dilation and erosion [21, 23, 24]. Randomization introduces stochastic components into the delineation of VOIs, mimicking realistic errors made when different radiologists delineate the tumors. The latter two can mimic more generous inclusion of background (dilation) or very stringent omission of background (erosion) by the person segmenting the tumours. Smoothing is comparable to an observer outlining the VOI with less attention to detail. Other modifications, such as ellipse fitting are an example of very simple and time efficient outlining of cancer as could be performed also by less well-trained readers.

We observed a relatively large reduction in prediction performance when randomization masks were used, confirming the high sensitivity of the radiomics models to random tumor delineation errors introduced by different radiologists. Furthermore, we found that the modifications of the original manual VOIs affect radiomics feature values, selection of features, and prediction performance. Even slight smoothing of VOIs affected which feature selection methods achieved the predictions with the highest AUC and which features were selected for the best prediction results. This is consistent with previously reported findings where tumor delineation also determined which features were selected [21, 23]. The reason is likely that VOI modifications, even though small, change radiomics feature values and thus result in a different order of features when ranked according to predictive power.

In TNBC, the predictive performance was more robust to VOI modifications than in HER2+ breast cancer because lower percentage changes in AUC observed for both train and test sets for TNBC regardless of whether only features selected based on the original manual VOIs were used or the feature selection process was repeated with each VOI modification.

The lowest robustness and predictive performance were found when using ellipsoid fitting and dilation 2, potentially due to relatively large amounts of normal breast tissue that is included in these ROIs. In contrast to a study on melanoma, where the inclusion of normal appearing skin surrounding the cancer contributed significantly to higher AUC of predictions [22], in the case of breast cancer the surrounding normal breast tissue apparently contributes little to the predictions. In contrast, higher robustness predictive performance were observed with smoothing and erosion 1. However, no similar relationship between robustness and prediction performance was found for other VOI modifications. This is also consistent with previous studies in other diseases where feature robustness and predictive performance were not necessarily intertwined when segmentations were modified [21, 23, 24]. Therefore, the joint analysis of both feature robustness and the final predictive performance of radiomics models need to be performed for a comprehensive evaluation of VOI differences and their effect on the development of prediction models [21, 54].

Stable and high predictive performance in VOIs with smoothing 1, smoothing 2, and erosion 1 for both cancer groups/types confirm the tolerance of the corresponding differences in radiomics models. This suggests that radiologists can smoothly outline the lesions with 1 mm or 2 mm or outline the tumor with 1 mm shrinkage. One of the limitations of this study is that no VOIs obtained from multiple readers were compared. However, the comparison of VOIs from different radiologists was beyond the scope of this study and can be found in the existing radiomics literature. This study set out to compare VOI modifications that were applied in a standardised way rather than observer-introduced variations. Furthermore, this study is limited to the evaluation of VOI modifications in breast cancer. In future studies, more types of cancer and different imaging modalities should be analyzed in order to extend the generalizability of the results.

5. **Conclusion**

Based on only one MRI sequence, we developed radiomics-based predictors of pCR in breast cancer. Our predictive performance with these limited data was similar to studies combining clinical and imaging data or radiomics from different MRI sequences. Our systematic evaluation showed that different VOI modifications can lead to significant differences in radiomics feature values, feature selection and prediction performance. Determining a predefined standard for tumor delineation can help develop reliable and robust radiomics models. The results of this study can serve as a reference for future radiomics research.